# Van der Waals equation of state and PVT – properties of real fluid

Umirzakov I. H.

Institute of Thermophysics, Pr. Lavrenteva St., 1, Novosibirsk, Russia, 630090
e-mail: umirzakov@itp.nsc.ru

**Abstract**

It is shown that: in the case when two parameters of the Van der Waals equation of state are defined from the critical temperature and pressure the exact parametrical solution of the equations of the liquid-vapor phase equilibrium of the Van der Waals fluid quantitatively describes the experimental dependencies of the saturated pressure of argon on the temperature and reduced vapor density; it can describe qualitatively the temperature dependencies of the reduced vapor and liquid densities of argon; and it gives the quantitative description of the temperature dependencies of the reduced densities near critical point. When the parameters are defined from the critical pressure and density the parametric solution describes quantitatively the experimental dependencies of the saturated pressure of argon on the density and reduced temperature, it can describe qualitatively the dependencies of the vapor and liquid densities on the reduced temperature, and it gives the quantitative description of the dependencies of the densities on the reduced temperature near critical point. If the parameters are defined from the critical temperature and density then the exact solution describes quantitatively the experimental dependencies of the reduced saturated pressure of argon on the density and temperature, it describes qualitatively the temperature dependencies of the vapor and liquid densities of argon, and it gives the quantitative description of the temperature dependencies of the vapor and liquid densities near critical point.

It is also shown that the Van der Waals equation of state describes quantitatively the reference experimental PVT- data for the gas and supercritical fluid states for the under-critical densities of argon, the dependencies of the saturation pressure on the temperature and vapor density, and the dependence of the vapor density of argon on temperature if the parameters are defined from the critical pressure and temperature.

**Keywords** Argon, Liquid-Vapor Phase Equilibrium, Fluid, Critical Point, Coexistence

**Introduction**

As known the similarity laws allow one to set a correspondence between different thermodynamic values without explicit use of the equation of state (EOS) [1]. Some of consequences of the van der Waals equation (VdW-EOS) are valid for a great number of various models and real systems described by completely different equations of state [1]. The well-known examples of such similarities are the principle of corresponding states and law of rectilinear diameter [2]. According to VdW-EOS an ideal line for pressure on temperature-density plane along which the compressibility factor is equal to unity is straight linear [3]. There is considerable experimental evidence, confirming the linearity of this line for many other substances and models [1, 2, 4-7]. This line is also named as "Zeno line" (obtained from "Z = one") [2]. This regularity appears to have much wider area of applicability than the original VDW equation [1]. Besides, the Zeno-line is used to check both the reference [8] and semiempirical [9] equations of states.

VdW-EOS is also used to generate new similarity relations [1, 10, 11]. Predictions of VdW-EOS for lines along which an enthalpy, free energy and chemical potential of van der Waals fluid coincides with their values for the ideal gas describe many substances and model systems [1, 6, 12, 13]. For more details see [1] and references here in.

There are two alternative theories of the critical region of the liquid-vapor first-order phase transitions. The first theory is the traditional theory of the region near single critical point based upon Ising-like scaling theory with crossover to classical equations of state [14-20]. VdW-EOS [18] and the fundamental equations of state [20], which are based upon the concept of a single gas-liquid critical point, are representations of these classical equations of state.

The second theory is the "meso-phase" hypothesis of Woodcock [26]. According to the "meso-phase" hypothesis: at critical and supercritical temperatures on the thermodynamic (density, pressure)-plane exists a region where the pure substance is in the "meso-phase", which consists of small clusters that are gas like and clusters of macroscopic size that are liquid like; there is exist a line of critical points over a finite range of densities at critical temperature and pressure instead of single critical point; and the pressure in the "meso-phase" is linear function of density. This hypothesis is reminiscent of an old concept of the supercritical fluid as a mixture of "gasons" and "liquidons" that has turned out to be inconsistent with the experimental evidence [16,17].

Some predictions of the "meso-phase" hypothesis were criticized by Sengers and Anisimov [16] and Umirzakov [27]. According to [16] in contrast to the conjecture of Woodcock, there is no reliable experimental evidence to doubt the existence of a single critical point in the thermodynamic limit and of the validity of the scaling theory for critical thermodynamic behavior.

According to [26] the Van der Waals critical point does not comply with the Gibbs phase rule and its existence is based upon a hypothesis rather than a thermodynamic definition. The paper [27] mathematically demonstrates that a critical point is not only based on a hypothesis that is used to define values of two parameters of VdW-EOS. Instead, the author of [27] argued that a critical point is a direct consequence of the thermodynamic phase equilibrium conditions resulting in a single critical point. It was shown that the thermodynamic conditions result in the first and second partial derivatives of pressure with respect to volume at constant temperature at a critical point equal to zero which are usual conditions of an existence of a critical point [27].

The papers [28] and [29] were the responses to the critique of some predictions of the hypothesis in [4] and [27], respectively.

The paper [28] was criticized in [30]. It was shown [30] that: (1) the expressions for the isochoric and isobaric ($C_P$) heat capacities of liquid and gas, coexisting in phase equilibrium, the heat capacities at saturation of liquid and gas ($C_\sigma$) and the heat capacity ($C_\lambda$) used in Woodcock's article [28] are incorrect; (2) the conclusions of the article based on the comparison of the incorrect $C_V$, $C_P$, $C_\sigma$ and $C_\lambda$ with experimental data are also incorrect; (3) the lever rule used in [28] cannot be used to define $C_V$ and $C_P$ in the two-phase coexistence region; and (4) the correct expression for the isochoric heat capacity describes the experimental data well.

The Van der Waals equation of state was criticized in [29]. According to [29] *"state functions of van der Waals' equation fail to describe the thermodynamic properties of gases and gas–liquid coexistence", "the van der Waals equation fails to describe even qualitatively the thermodynamic properties of gas–liquid coexistence in the critical region, and "the liquid–gas critical point is not a property that the van der Waals equation can make any statements about".*

We show in this paper that above statements of [29] are incorrect. We show that in the case when two parameters of Van der Waals equation of state are defined from the critical temperature and pressure the exact parametrical solution of the equations of the liquid-vapor phase equilibrium of Van der Waals fluid (VdW-fluid), which is described by VdW-EOS, quantitatively describes the experimental dependencies of the saturated pressure of argon on the temperature and reduced vapor density; it can describe qualitatively the temperature dependencies of the reduced vapor and liquid densities of argon; and it gives the quantitative description of the temperature dependencies of the reduced densities near critical point. When the parameters are defined from the critical pressure and density, the parametric solution describes quantitatively the experimental dependencies of the saturation pressure of argon on the density and reduced temperature, it can describe qualitatively the dependencies of the vapor and liquid densities on the reduced temperature, and it gives the quantitative description of the dependencies of the densities on the reduced temperature near critical point. If the parameters are defined from the critical temperature and density then the exact solution describes quantitatively the experimental dependencies of the reduced saturated pressure of argon on density and temperature, it describes qualitatively the temperature dependencies of the vapor and liquid densities of argon, and it gives the quantitative description of the temperature dependencies of the vapor and liquid densities near critical point. It is also shown that the Van der Waals equation of state describes quantitatively the experimental dependencies of the saturation pressure on the temperature and vapor density of argon [22], and the experimental dependence of the vapor density of argon on temperature if the parameters are defined from the critical pressure and temperature, and VDW-EOS describes quantitatively the reference experimental PVT- data [21] for gas state and supercritical fluid states for under-critical densities of argon.

**Comparison of parametric solution of the coexistence equations of Van der Waals fluid with liquid-vapor equilibrium of real fluid**

The Van der Waals' equation of state (VdW-EOS) [18] is the relation

$$p(n,T) = nkT/(1-bn) - an^2, \qquad (1)$$

where $p$ is the pressure, $T$ is the temperature, $n$ is the number density of the particles (atoms or molecules), $k$ is the Boltzmann constant, $a$ and $b$ are the positive constants. The values of the number density, pressure and temperature at the critical point of liquid-vapor first-order phase transition are equal to $n_c = 1/3b$, $p_c = a/27b^2$ and $T_c = 8a/27kb$, respectively [18, 23-25]. The mass density $\rho$ is equal to $\rho = n\mu$, where $\mu$ is the mass of the particle.

It is easy to see that VdW-EOS defines the exact position of the critical point on the thermodynamic ($p,T$)- , ($n,p$)- and ($T,n$)- planes and, hence, it describes the density-temperature-pressure relation and phase equilibrium line near the critical point (see Figs. 1-3) if the coefficients $a$ and $b$ of VdW-EOS are defined from ($p_c,T_c$), ($p_c,n_c$) and ($T_c,n_c$) using the relations

$$a = 27k^2T_c^2/64p_c, \qquad b = kT_c/8p_c, \qquad (2)$$
$$a = 3p_c/n_c^2, \qquad b = 1/3n_c, \qquad (3)$$
$$a = 9kT_c/8n_c, \qquad b = 1/3n_c, \qquad (4)$$

respectively. Here $p_c$, $T_c$ and $n_c$ are the values of the pressure, temperature and density of the real fluid at the critical point.

According to the parametric solution [23-25, 27] of the equations corresponding to the liquid-vapor phase equilibrium of VdW-fluid the saturation pressure $p_e(T)$ and the densities of liquid $n_L(T)$ and vapor $n_V(T)$ of VdW-fluid are defined

$$bkT/a = F(y(T)) \equiv \frac{2(y - e^{2y} + ye^{2y} + 1)(4ye^{2y} - e^{4y} + 1)^2}{(e^{2y} - 1)(2y - 2e^{2y} + e^{4y} - 2ye^{4y} + 4y^2 e^{2y} + 1)^2}\bigg|_{y=y(T)}, \quad (5)$$

$$bn_L(T) = F_L(y(T)) = 2 \cdot \frac{y - e^{2y} + ye^{2y} + 1}{(2y - 1)e^{2y} - e^{-2y} - 2y + 2}\bigg|_{y=y(T)}, \quad (6)$$

$$bn_V(T) = F_V(y(T)) \equiv 2 \cdot \frac{y - e^{2y} + ye^{2y} + 1}{e^{4y} - e^{2y}(2y+2) + 2y + 1}\bigg|_{y=y(T)}, \quad (7)$$

$$p_e(T)b^2/a = F_P(y(T)) \equiv \left(F(y)F_L(y)/[1 - F_L(y)] - F_L^2(y)\right)\bigg|_{y=y(T)}. \quad (8)$$

The temperature dependencies of the saturation pressure $p_e(T)$ and the densities of liquid $n_L(T)$ and vapor $n_V(T)$ of VdW-fluid are defined from Eqs. 6-8 using the temperature dependence of the parameter $y(T)$ defined from Eq. 5.

We have from Eqs. 5-8 using Eq. 2

$$T = 27T_c/8 \cdot F(y(T)), \quad (9)$$

$$n_L(T)/n_c = 8z_c F_L(y(T)), \quad (10)$$

$$n_V(T)/n_c = 8z_c F_V(y(T)), \quad (11)$$

$$p_e(T) = 27 p_c F_P(y(T)). \quad (12)$$

where $z_c = p_c/n_c kT_c$ is the value of the compressibility factor of the real fluid at the critical point. The dependencies $p_e(T)$, $n_L(T)$ and $n_V(T)$ are defined from Eqs. 10-12 using the temperature dependence of the parameter $y(T)$ defined from Eq. 9.

Fig. 1 demonstrates that Eqs. 9-12 - the exact parametrical solution of the equations of the liquid-vapor phase equilibrium of VdW-fluid - quantitatively describe the experimental dependencies [22] of the saturated pressure of argon on the temperature and reduced (to critical) vapor density near critical point, and they can describe qualitatively the reduced vapor and liquid densities of argon on the temperature in the case when the parameters $a$ and $b$ are defined from $(T_c, p_c)$ using Eq. 2 and $T_c = 150.567\,K$, $\mu = 0.039948\,kg \cdot mol^{-1}$, $\rho_c = 535.6\,kg \cdot m^{-3}$ and $p_c = 4.863\,MPa$ for argon [21,22].

VdW-EOS predicts in this case for the critical mass density $418\,kg \cdot m^{-3}$ which is considerably less than $\rho_c = 535.6\,kg \cdot m^{-3}$ of argon [22]. Therefore VdW-EOS describes the saturation pressure as the function of the reduced vapor density (Fig. 1b) and the reduced liquid and vapor densities as the function of temperature (Fig. 1c).

We have from Eqs. 5-8 using Eq. 3

$$T/T_c = 1/9z_c \cdot F(y(T)), \quad (13)$$

$$n_L(T) = 3n_c F_L(y(T)), \quad (14)$$

$$n_V(T) = 3n_c F_V(y(T)), \quad (15)$$

$$p_e(T) = 27 p_c F_P(y(T)).  \qquad (16)$$

The dependencies $p_e(T)$, $n_L(T)$ and $n_V(T)$ are defined from Eqs. 14-16 using the temperature dependence of the parameter $y(T)$ defined from Eq. 13.

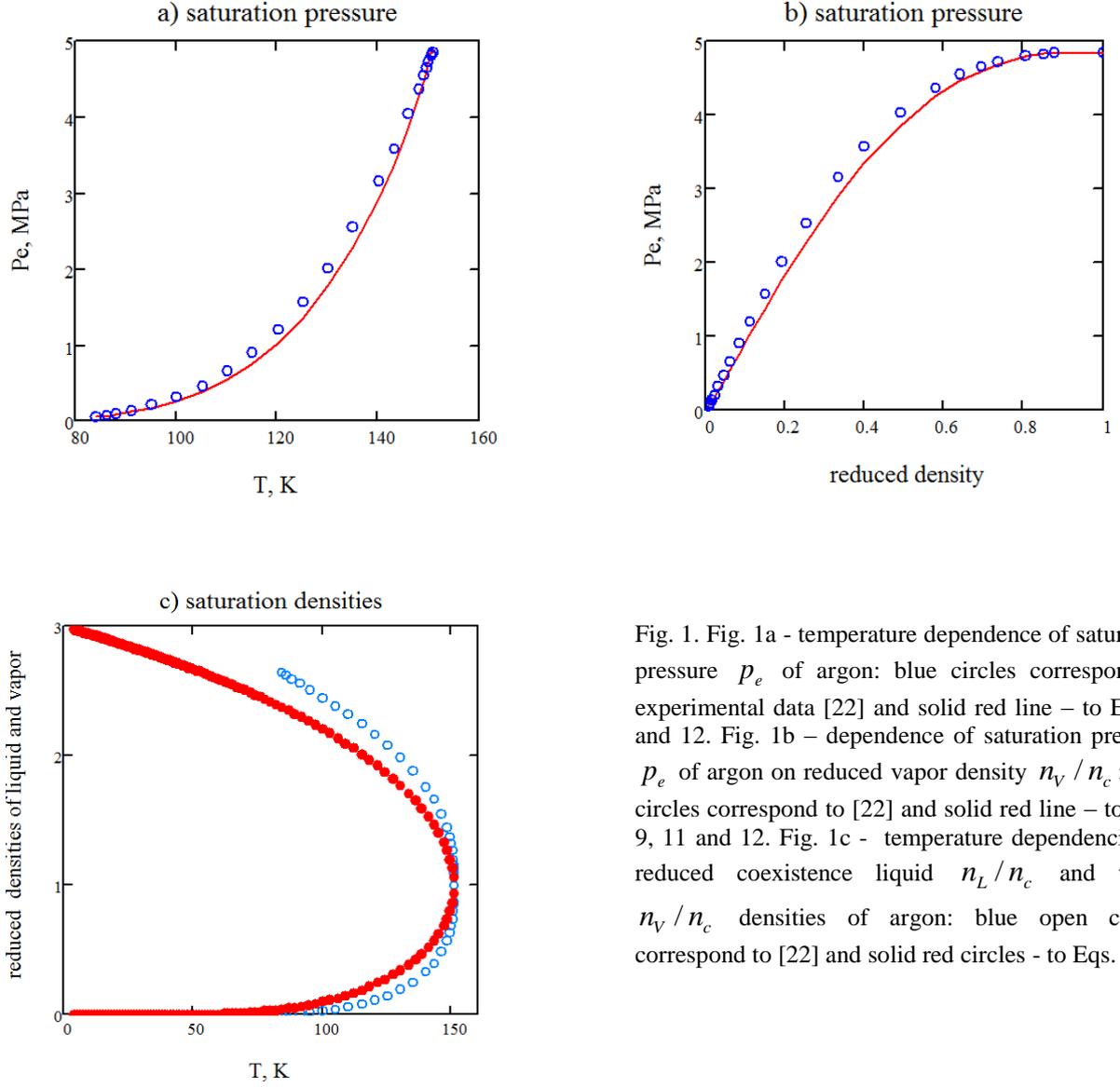

Fig. 1. Fig. 1a - temperature dependence of saturation pressure $p_e$ of argon: blue circles correspond to experimental data [22] and solid red line – to Eqs. 9 and 12. Fig. 1b – dependence of saturation pressure $p_e$ of argon on reduced vapor density $n_V / n_c$: blue circles correspond to [22] and solid red line – to Eqs. 9, 11 and 12. Fig. 1c - temperature dependencies of reduced coexistence liquid $n_L / n_c$ and vapor $n_V / n_c$ densities of argon: blue open circles correspond to [22] and solid red circles - to Eqs. 9-11.

Fig. 2 shows that Eqs. 13-16 - the exact parametrical solution of the equations of the liquid-vapor phase equilibrium of VdW-fluid describes quantitatively the experimental dependencies [22] of the saturated pressure of argon on the density and reduced (to critical) temperature near critical point, and it can describe qualitatively the vapor and liquid densities of argon on the reduced temperature when the parameters $a$ and $b$ are defined from $(n_c, p_c)$ using Eq. 3. VdW-EOS predicts in this case for the critical temperature $350\,K$ which is considerably greater than $T_c = 150.687\,K$ of argon [22]. Therefore VdW-EOS describes the saturation pressure (Fig. 2a) and the liquid and vapor densities (Fig. 2c) as the functions of the reduced temperature.

We have from Eqs. 5-8 using Eq. 4

$$T = 27 T_c / 8 \cdot F(y(T)),  \qquad (17)$$

$$n_L(T) = 3n_c F_L(y(T)),  \qquad (18)$$
$$n_V(T) = 3n_c F_V(y(T)),  \qquad (19)$$
$$p_e(T)/p_c = 81/8z_c \cdot F_P(y(T)). \qquad (20)$$

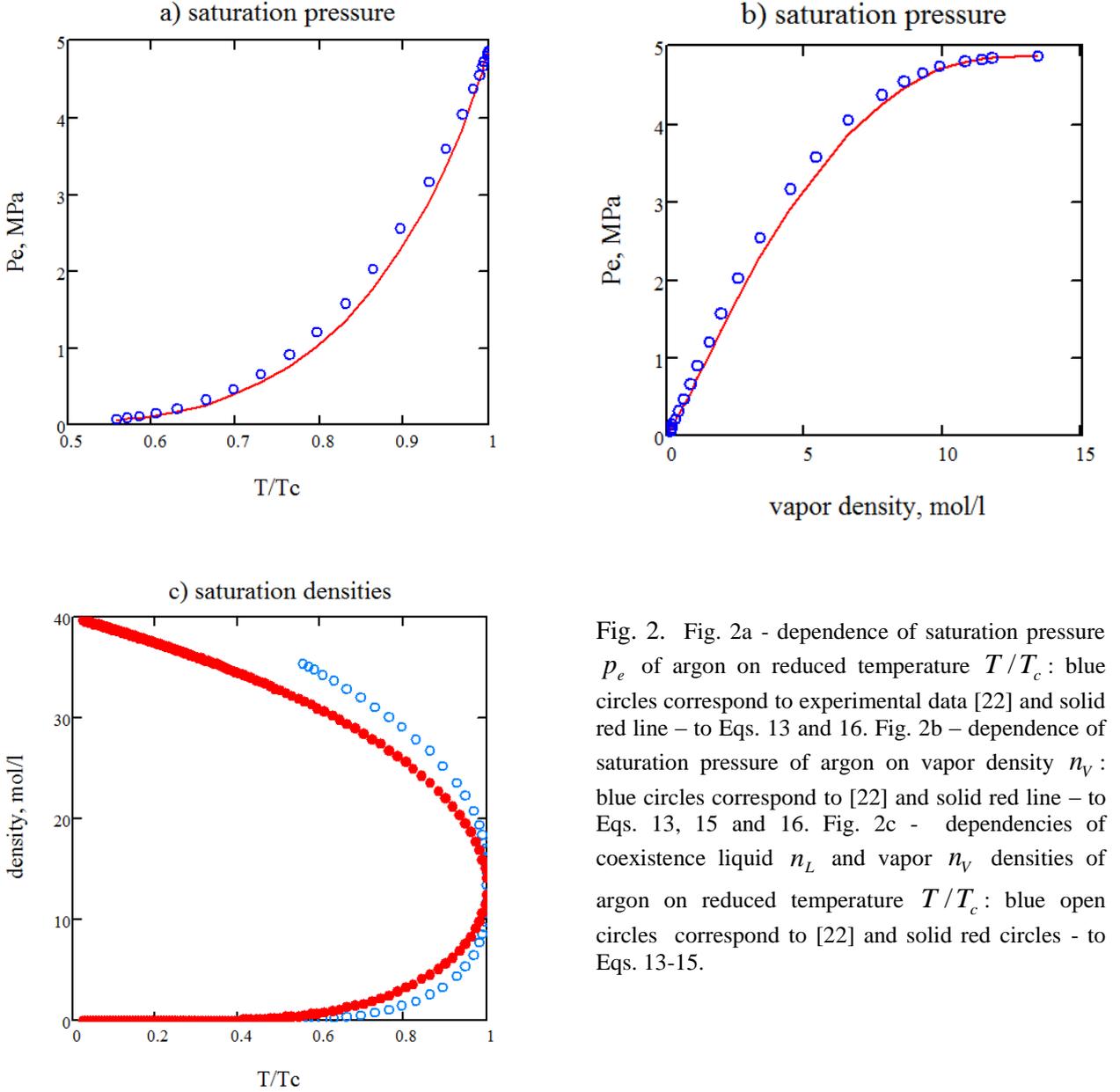

Fig. 2. Fig. 2a - dependence of saturation pressure $p_e$ of argon on reduced temperature $T/T_c$: blue circles correspond to experimental data [22] and solid red line – to Eqs. 13 and 16. Fig. 2b – dependence of saturation pressure of argon on vapor density $n_V$: blue circles correspond to [22] and solid red line – to Eqs. 13, 15 and 16. Fig. 2c - dependencies of coexistence liquid $n_L$ and vapor $n_V$ densities of argon on reduced temperature $T/T_c$: blue open circles correspond to [22] and solid red circles - to Eqs. 13-15.

The dependencies $p_e(T)$, $n_L(T)$ and $n_V(T)$ are defined from Eqs. 18-20 using the temperature dependence of the parameter $y(T)$ defined from Eq. 17.

Fig. 3 shows that Eqs. 17-20 describe quantitatively the experimental dependencies [22] of the reduced (to critical) saturated pressure of argon on density and temperature near critical point, and it describes qualitatively the vapor and liquid densities of argon on temperature when the parameters $a$ and $b$ are defined from $(n_c, T_c)$ using Eq. 4. VdW-EOS predicts in this case for the critical pressure $6.27\,MPa$ which is considerable greater than $p_c = 4.863\,MPa$ of argon [22]. Therefore VdW-EOS describes the reduced saturation pressure as the functions of the temperature (Fig. 3a) and vapor pressure (Fig. 3b).

One can see from Eqs. 5-6 [27] and Fig. 1 [27] that the saturated vapor density of the VdW-fluid is non-negative for any value of the temperature. Figs. 1c, 2c and 3c show the same. So, in contrast to the Capture of Fig. 4 [29], VdW-EOS is not absurd for temperatures below 110 $K$.

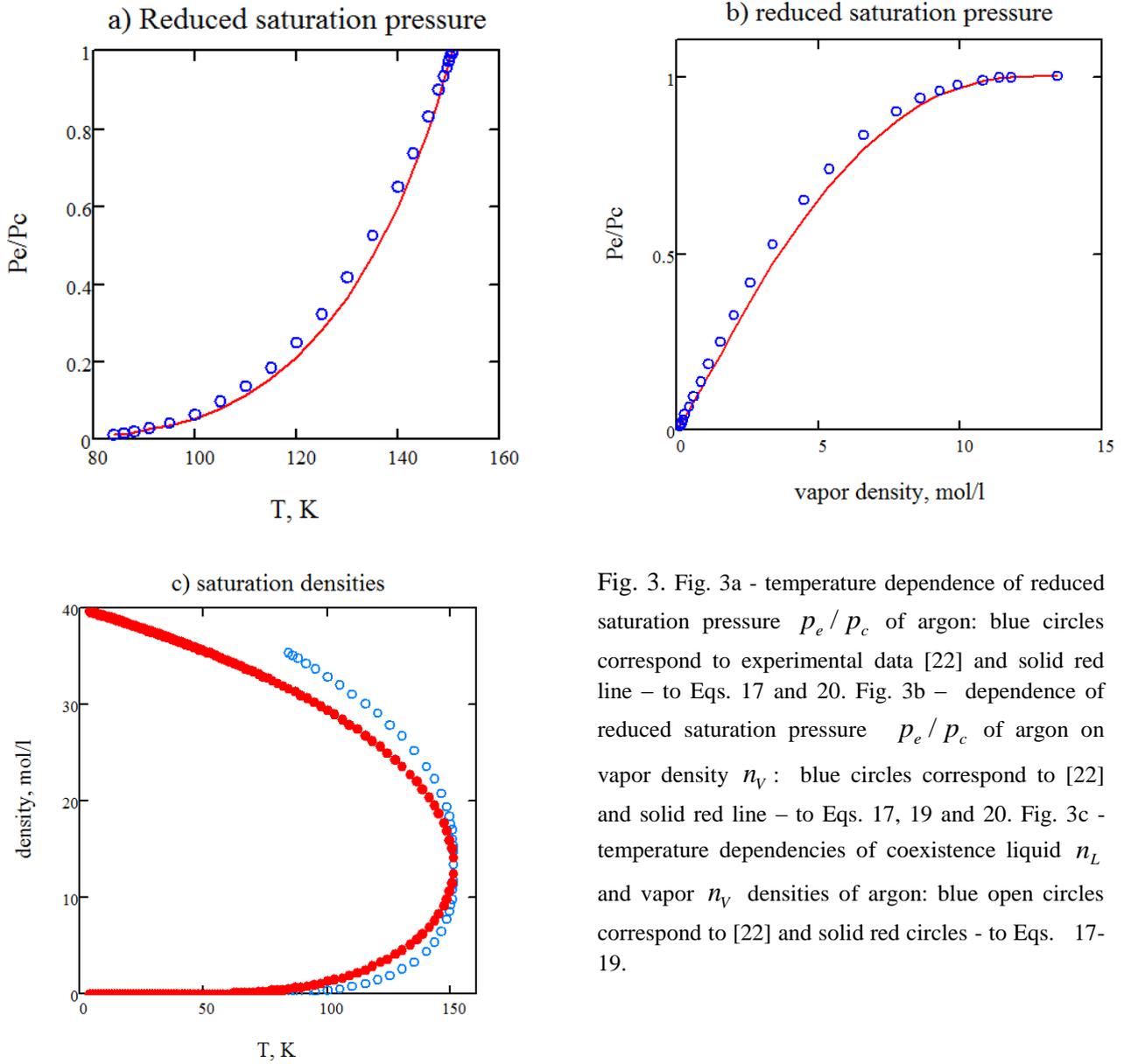

Fig. 3. Fig. 3a - temperature dependence of reduced saturation pressure $p_e/p_c$ of argon: blue circles correspond to experimental data [22] and solid red line – to Eqs. 17 and 20. Fig. 3b – dependence of reduced saturation pressure $p_e/p_c$ of argon on vapor density $n_V$: blue circles correspond to [22] and solid red line – to Eqs. 17, 19 and 20. Fig. 3c - temperature dependencies of coexistence liquid $n_L$ and vapor $n_V$ densities of argon: blue open circles correspond to [22] and solid red circles - to Eqs. 17-19.

**Van der Waals equation of state and liquid-vapor equilibrium of real fluid**

We obtain from Eq. 1 using Eqs. 2-4

$$p(n,T) = \frac{8p_c nkT}{8p_c - nkT_c} - \frac{27k^2 T_c^2 n^2}{64 p_c}, \tag{21}$$

$$p(n,T) = \frac{3nkT}{3 - n/n_c} - \frac{3p_c n^2}{n_c^2}, \tag{22}$$

$$p(n,T) = \frac{3nkT}{3 - n/n_c} - \frac{9kT_c n^2}{8n_c}. \tag{23}$$

It is easy to see from Eqs. 21-23 that one can define the temperature as the function of the pressure and density by use of the following equations

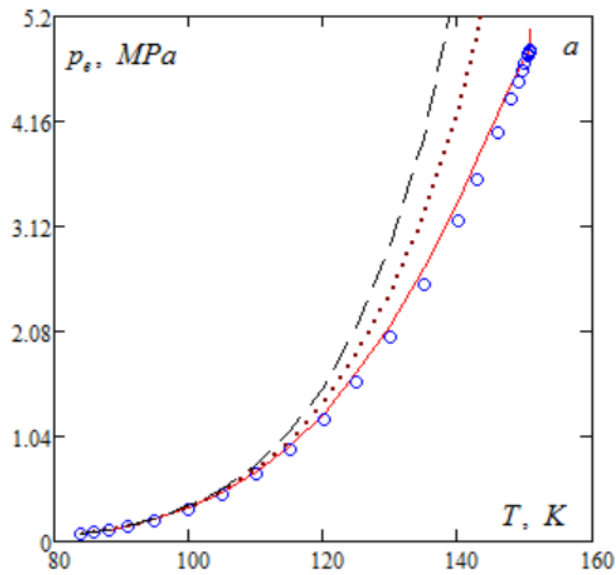
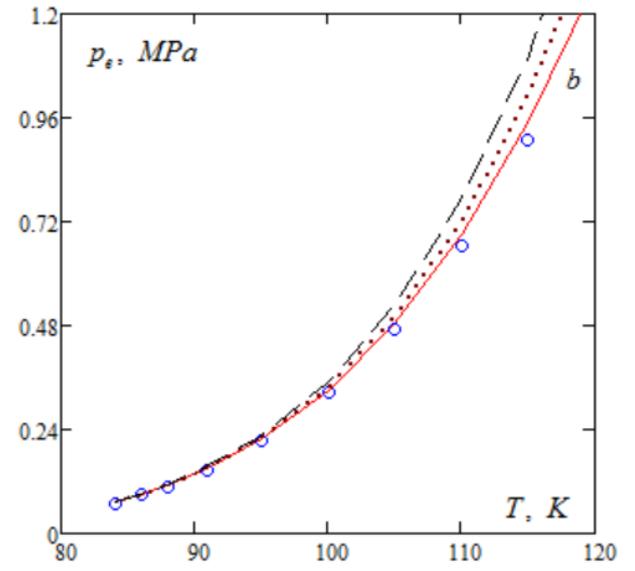
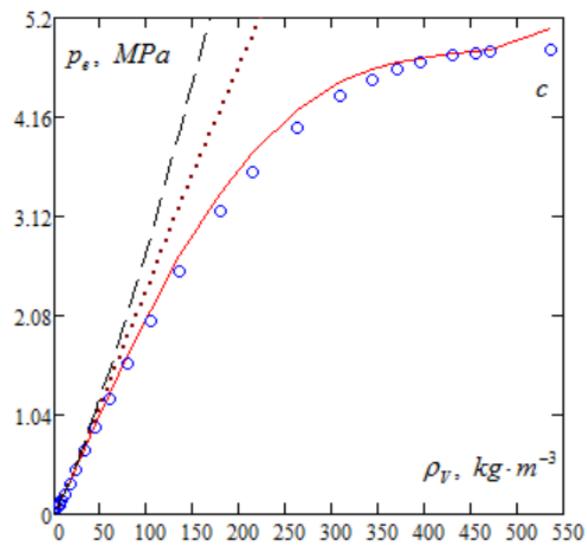
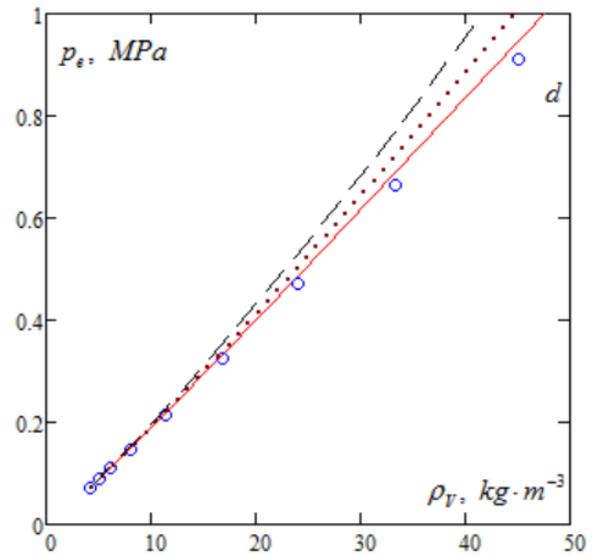

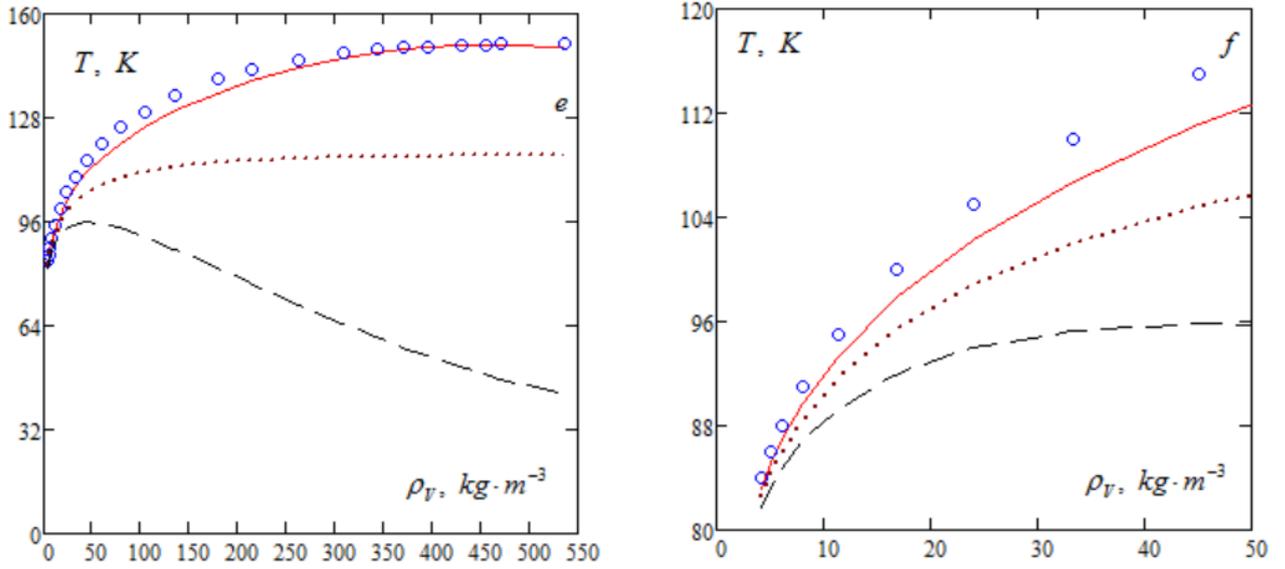

Fig. 4. Blue circles correspond to experimental data [22]. Figs. 4a, 4b - temperature dependence of saturation pressure $p_e$ of argon: solid red line – to Eq. 21, dotted brown line – to Eq. 22 and dashed black line to Eq. 23. Figs.4c, 4d – dependence of saturation pressure $p_e$ of argon on vapor mass density $\rho_V$: blue circles correspond to [22], solid red line – to Eq. 21, dotted brown line – to Eq. 22 and dashed black line to Eq. 23. Fig. 4e, 4f - dependence of temperature on saturated vapor mass density of argon $\rho_V$: solid red line – to Eq. 24, dotted brown line – to Eq. 25 and dashed black line to Eq. 26.

$$T(n,p) = \frac{1}{nk}\left(p + \frac{27k^2T_c^2n^2}{64p_c}\right)\cdot\left(1 - \frac{nkT_c}{8p_c}\right), \qquad (24)$$

$$T(n,p) = \frac{1}{3nk}\left(p + p_c\frac{3n^2}{n_c^2}\right)\cdot\left(3 - \frac{n}{n_c}\right), \qquad (25)$$

$$T(n,p) = \frac{1}{3nk}\left(p + \frac{9kT_cn^2}{8n_c}\right)\cdot\left(3 - \frac{n}{n_c}\right). \qquad (26)$$

The dependence of the saturated vapor pressure of argon on the temperature is shown on Figs. 4a and 4b. As one can see Eq. 21 describes quantitatively the experimental data [22] in the interval from the triple point temperature to the critical one, and Eqs. 22-23 describe quantitatively the data near triple point temperature only. Fig. 4b shows that the Eq. 21 gives the best description of the data [22] on the saturated vapor pressure near triple point temperature.

The dependence of the saturated vapor pressure of argon on the mass density of the saturated vapor is presented on Figs. 4c and 4d. As one can see Eq. 21 describes quantitatively the experimental data [22] in the interval from the triple point temperature to the critical one, and Eqs. 22-23 describe quantitatively the data near triple point temperature. Fig. 4d demonstrates that Eq. 21 gives the best description of the data [22] near triple point.

The dependence of the temperature on the saturated vapor mass density of argon $\rho_V = n_V\mu$ is presented on Figs. 4e and 4f. As evident Eq. 24 describes quantitatively the experimental data [22] in the interval from the triple point temperature to the critical one, and Eqs. 25-26 describe quantitatively the data near triple point temperature. Fig. 4f shows that the Eq. 24 gives the best description of the data [22] near triple point temperature.

# Van der Waals equation of state and PVT-properties of real fluid

The comparison of the Van der Waals equation of state (Eq. 21) with the reference experimental PVT–data [21] is shown on Figs. 5. The data for the vapor density on the coexistence line [22] are also included to the comparison. The values of temperatures corresponding to isotherms of [21, 22] (column 1), the phase state of argon (column 2, where "scf" denotes the supercritical state), the density region of the experimental data [21] described by VdW-EOS (column 3), number of experimental points of [21, 22] described by VdW-EOS (column 4), the full number of experimental points at the isotherm (column 5), and the value of the quadratic root of the mean of the sum of the squares of the displacements (square root of the standard deviation)

$$\Delta = \left[ \frac{1}{N-1} \sum_{i=1}^{N} \left( \frac{8 p_c n_i k T_i}{8 p_c - n_i k T_c} - \frac{27 k^2 T_c^2 n_i^2}{64 p_c} - p_i \right) \right]^{1/2} \cdot 100\%$$

of the predictions of VdW-EOS from the data [21,22] (column 6) for the isotherm are presented in Table 1. The abbreviation "oe" ("ue") in the column 7 of Table 1 denotes that VdW-EOS overestimates (underestimates) the data on the isotherm.

Table 1 and Fig. 5 show that Eq. 21 describes nine experimental isotherms of the gas state of argon and its eighteen supercritical experimental isotherms [21, 22] for the under-critical densities with the good accuracy. As one can see from the last column of Table 1 VDW-EOS (Eq. 21) overestimates all nine isotherms for the temperatures that are lesser or equal to the critical one, and it underestimates ten isotherms from the temperature interval $190\,K \leq T \leq 340\,K$. One can conclude using the data from the last row of Table 1 that Eq. 21 describes with a good accuracy 78% of the experimental data [21, 22] for the gas and supercritical states of argon for the densities less than the critical one. The full number of experimental PVT-data is equal to 638 [21] therefore Eq. 21 describes quantitatively 66% of them. The columns 4 and 5 show that Eq. 21 describes quantitatively all experimental PVT-data of gas at 8 isotherms from the temperature interval $110\,K \leq T \leq 148\,K$ and all data at 10 supercritical isotherms from the temperature interval $190\,K \leq T \leq 340\,K$.

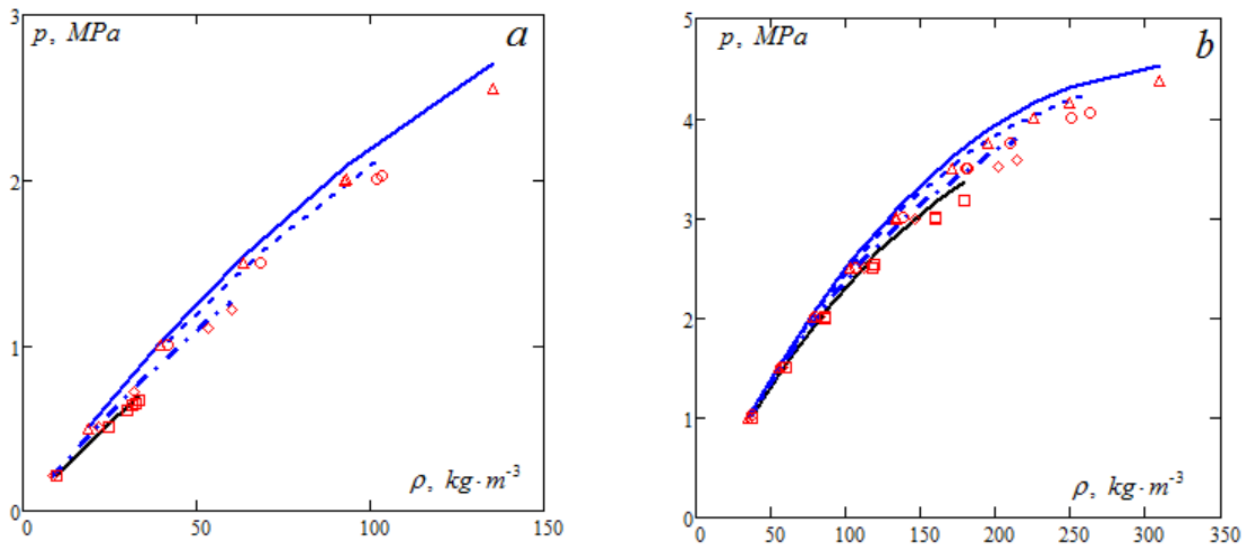

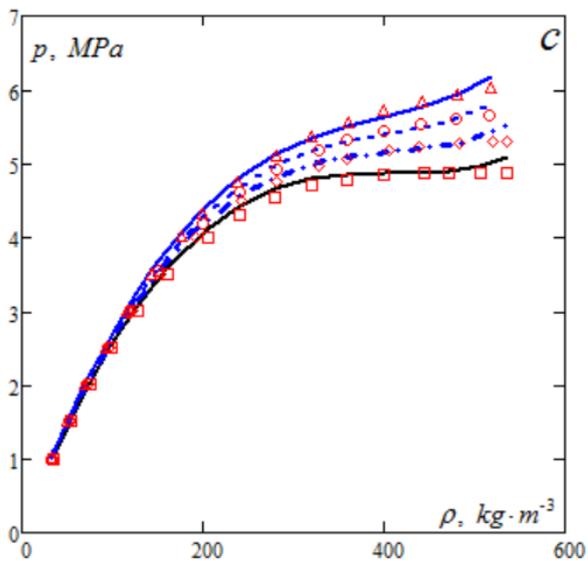
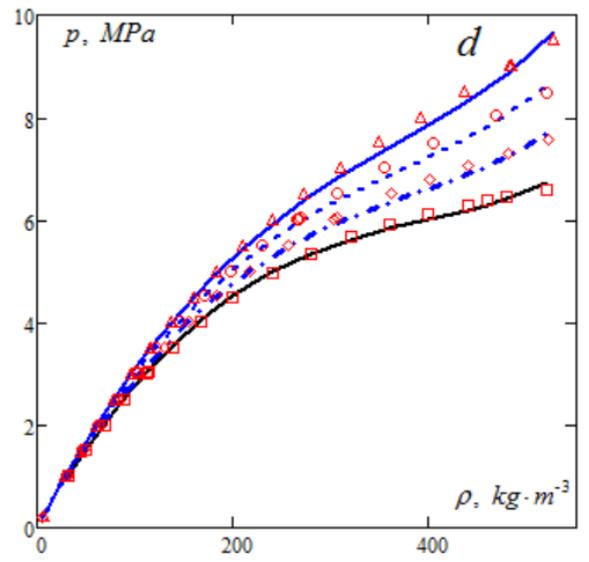
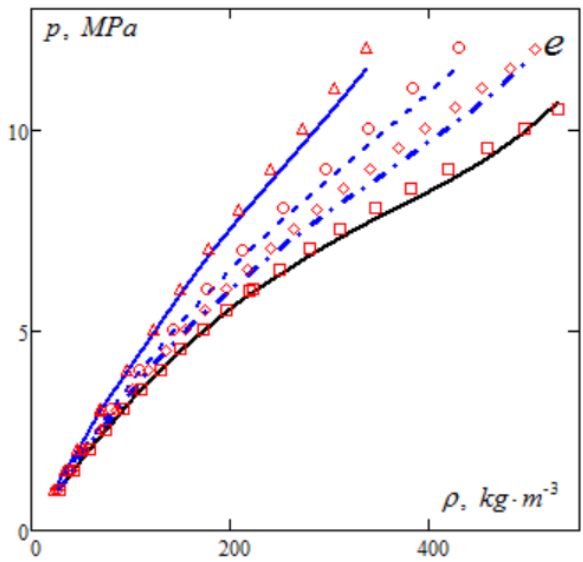
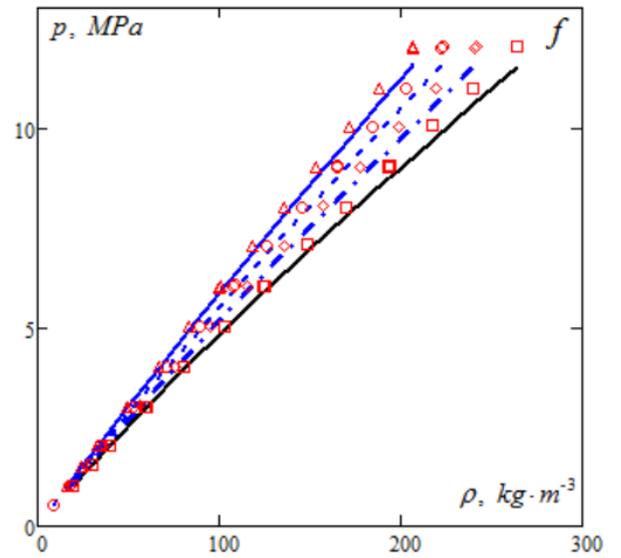
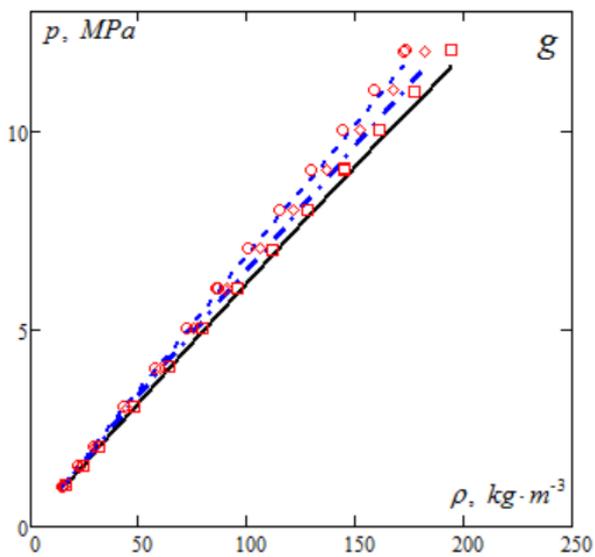

Fig. 5. Isotherms of argon: red symbols correspond to experimental data [21] and lines – to VDW-EOS Eq. 21. a) Black solid line and squares correspond to $110\,K$, blue dotted-dashed line and diamonds – to $120\,K$, blue dashed line and circles – to $130\,K$, and blue solid line and triangles – to $135\,K$. b) Black solid line and squares correspond to $140\,K$, blue dotted-dashed line and diamonds – to $143\,K$, blue dashed line and circles – to $146\,K$, and blue solid line and triangles – to $148\,K$. c) Black solid line and squares correspond to $150.7\,K$, blue dotted-dashed line and diamonds – to $153\,K$, blue dashed line and circles – to $155\,K$, and blue solid line and triangles – to $157\,K$. d) Black solid line and squares correspond to $160\,K$, blue dotted-dashed line and diamonds – to $165\,K$, blue dashed line and circles – to $170\,K$, and blue solid line and triangles – to $175\,K$. e) Black solid line and squares correspond to $180\,K$, blue dotted-dashed line and diamonds – to $190\,K$, blue dashed line and circles – to $200\,K$, and blue solid line and triangles – to $220\,K$. f) Black solid line and squares correspond to $250\,K$, blue dotted-dashed line and diamonds – to $265\,K$, blue dashed line and circles – to $280\,K$, and blue solid line and triangles – to $295\,K$. g) Black solid line and squares correspond to $310\,K$, blue dotted-dashed line and diamonds – to $325\,K$, and blue dashed line and circles – to $340\,K$.

**Table 1. Description of experimental isotherms of argon [21] presented on Fig. 5.**

| 1 | 2 | 3 | 4 | 5 | 6 | 7 |
|---|---|---|---|---|---|---|
| $T,\,K$ | phase state of argon | density region | number of described points | full number of exp.points | $\Delta$, % | oe - overestimates ue - underestimates |
| 110 | gas | $\rho \leq \rho_V(T)$ | 6 | 6 | 3.2 | oe |
| 120 | gas | $\rho \leq \rho_V(T)$ | 6 | 6 | 3.1 | oe |
| 130 | gas | $\rho \leq \rho_V(T)$ | 5 | 5 | 4.7 | oe |
| 135 | gas | $\rho \leq \rho_V(T)$ | 7 | 7 | 4.0 | oe |
| 140 | gas | $\rho \leq \rho_V(T)$ | 9 | 9 | 4.3 | oe |
| 143 | gas | $\rho \leq \rho_V(T)$ | 9 | 9 | 3.7 | oe |
| 146 | gas | $\rho \leq \rho_V(T)$ | 10 | 10 | 3.7 | oe |
| 148 | gas | $\rho \leq \rho_V(T)$ | 13 | 13 | 2.8 | oe |
| 150.7 | gas | $\rho \leq \rho_c$ | 16 | 33 | 2.2 | oe |
| 153 | scf | $\rho < \rho_c$ | 17 | 41 | 1.9 | |
| 155 | scf | $\rho < \rho_c$ | 17 | 34 | 1.3 | |
| 157 | scf | $\rho < \rho_c$ | 16 | 31 | 1.3 | |
| 160 | scf | $\rho < \rho_c$ | 19 | 33 | 1.2 | |
| 165 | scf | $\rho < \rho_c$ | 19 | 31 | 1.2 | |
| 170 | scf | $\rho < \rho_c$ | 19 | 28 | 1.3 | |
| 175 | scf | $\rho < \rho_c$ | 22 | 28 | 1.6 | |
| 180 | scf | $\rho < \rho_c$ | 23 | 27 | 1.9 | |
| 190 | scf | $\rho < \rho_c$ | 27 | 27 | 2.3 | ue |
| 200 | scf | $\rho < \rho_c$ | 17 | 17 | 2.7 | ue |
| 220 | scf | $\rho < \rho_c$ | 17 | 17 | 2.8 | ue |
| 250 | scf | $\rho < \rho_c$ | 18 | 18 | 2.7 | ue |
| 265 | scf | $\rho < \rho_c$ | 17 | 17 | 2.5 | ue |
| 280 | scf | $\rho < \rho_c$ | 20 | 20 | 2.5 | ue |
| 295 | scf | $\rho < \rho_c$ | 17 | 17 | 2.3 | ue |
| 310 | scf | $\rho < \rho_c$ | 17 | 17 | 2.2 | ue |
| 325 | scf | $\rho < \rho_c$ | 20 | 20 | 1.9 | ue |
| 340 | scf | $\rho < \rho_c$ | 17 | 17 | 2.0 | ue |
| | | | Σ420 | Σ538 | | |

**Conclusion**

We showed that:

1) in the case when two parameters of Van der Waals equation of state are defined from the critical temperature and pressure the exact parametrical solution of the equations of the liquid-vapor phase equilibrium of Van der Waals fluid quantitatively describes the experimental dependencies of the saturated pressure of argon on the temperature and reduced vapor density; it can describe qualitatively the temperature dependencies of the reduced vapor and liquid densities of argon;  and it gives the quantitative description of the temperature dependencies of the reduced densities near critical point;
2) when the parameters are defined from the critical pressure and density, the parametric solution describes quantitatively the experimental dependencies of the saturated pressure of argon on the density and reduced temperature, it can describe qualitatively the dependencies of the vapor and liquid densities on the reduced temperature, and it gives the quantitative description of the dependencies of the densities on the reduced temperature near critical point;
3) if the parameters are defined from the critical temperature and density then the exact solution describes quantitatively the experimental dependencies of the reduced saturated pressure of argon on the density and temperature, it describes qualitatively the temperature dependencies of the vapor and liquid densities of argon, and it gives the quantitative description of the temperature dependencies of the vapor and liquid densities near critical point;
4) the Van der Waals equation of state describes quantitatively the experimental dependencies of the saturation pressure on the temperature and vapor density [22], and the experimental dependence of the vapor density of argon on the temperature if the parameters are defined from the critical pressure and temperature;
5) the Van der Waals equation of state describes quantitatively the reference experimental PVT- data [21] for gas and supercritical fluid states for the  under-critical densities of argon if the parameters are defined from the critical pressure and temperature .